**Defect engineering of silicon with ion pulses from laser acceleration**

Walid Redjem[1], Ariel J. Amsellem[2], Frances I. Allen[3, 4], Gabriele Benndorf[5], Jianhui Bin[2,6], Stepan Bulanov[2], Eric Esarey[2], Leonard C. Feldman[7], Javier Ferrer Fernandez[8], Javier Garcia Lopez[8], Laura Geulig[2, 10], Cameron R. Geddes[2], Hussein Hijazi[7], Qing Ji[2], Vsevolod Ivanov[2,9], Boubacar Kante[1,2], Anthony Gonsalves[2], Jan Meijer[5], Kei Nakamura[2], Arun Persaud[2], Ian Pong[2], Lieselotte Obst-Huebl[2], Peter A. Seidl[2], Jacopo Simoni[9], Carl Schroeder[2,11], Sven Steinke[2,12], Liang Z. Tan, Ralf Wunderlich[5], Brian Wynne[2], and Thomas Schenkel[2,*]

[1]Department of Electrical Engineering and Computer Science, University of California, Berkeley, CA 94720, USA

[2]Accelerator Technology and Applied Physics Division, Lawrence Berkeley National Laboratory, Berkeley, CA 94720, USA

[3]Department of Materials Science and Engineering, University of California, Berkeley, CA 94720, USA

[4]California Institute for Quantitative Biosciences, University of California, Berkeley, CA 94720, USA

[5]Felix-Bloch Institute for Solid State Physics, Department: Applied Quantum Systems, Linnéstr. 5, D-04103 Leipzig, Germany

[6]present address: State Key Laboratory of High Field Laser Physics and CAS Center for Excellence in Ultra-Intense Laser Science, Shanghai Institute of Optics and Fine Mechanics, Chinese Academy of Sciences, Shanghai 201800, China

[7]Rutgers University, Department of Physics & Astronomy, 136 Frelinghuysen Rd, Piscataway, NJ 08854, USA

[8]University of Seville, Seville, Facultad de Física, Centro Nacional de Aceleradores, 41092 Sevilla, Spain

[9]Molecular Foundry, Lawrence Berkeley National Laboratory, Berkeley, CA 94720, USA

[10]present address: Fakultät für Physik, Ludwig-Maximilians-Universität München, Am Coulombwall 1, 85748 Garching, Germany

[11]Department of Nuclear Engineering, University of California, Berkeley, CA 94720, USA

[12]Present address: Marvel Fusion GmbH, Blumenstrasse 28, D-80331 Munich, Germany

*corresponding author: T_Schenkel@LBL.gov

**Abstract**

Defect engineering is foundational to classical electronic device development and for emerging quantum devices. Here, we report on defect engineering of silicon single crystals with ion pulses from a laser accelerator with ion flux levels up to $10^{22}$ ions/cm$^2$/s. Low energy ions from plasma expansion of the laser-foil target are implanted near the surface and then diffuse into silicon samples that were locally pre-heated by high energy ions. We observe low energy ion fluences of ~$10^{16}$ cm$^{-2}$, about four orders of magnitude higher than the fluence of high energy (MeV) ions. In the areas of highest energy deposition, silicon crystals exfoliate from single ion pulses. Color centers, predominantly W and G-centers, form directly in response to ion pulses without a subsequent annealing step. We find that the linewidths of G-centers increase in areas with high ion flux much more than the linewidth of W-centers, consistent with density functional theory calculations of their electronic structure. Laser ion acceleration generates aligned pulses of high and low energy ions that expand the parameter range for defect engineering and doping of semiconductors with tunable balances of ion flux, damage rates and local heating.

Control of the electronic structure of semiconductors through doping and defect engineering has enabled large-scale integration of classical electronic devices[1]. Ion implantation emerged in the 1960s as an effective and economically viable method to introduce dopant atoms into semiconductors such as silicon [2]. In a typical process flow, implantation of ions with energies from ~1 keV to 1 MeV is followed by a thermal annealing step to repair damage to the host crystal lattice that was induced during the stopping of energetic ions, to activate dopants, or to form desired defects [3][4]. A prominent example of a defect for emerging applications in quantum information science is the nitrogen-vacancy color center in diamond that can be formed by implantation of nitrogen ions followed by thermal annealing [5]. Color centers and photon emitting defects are also ubiquitous in other semiconductors, including in silicon, and have long been known to form during exposure of silicon crystals to high energy particles and photons [6]. Recently, photon emitting defect centers in silicon, including the G-center (a pair of substitutional carbon atoms bound to a silicon interstitial atom) and the W-center (three silicon interstitial atoms), have been revisited for quantum applications with single photon sources [7][8][9][10]. The promise of compatibility with CMOS processing for (ultra) large scale integration makes this particularly intriguing [10,11]. The reliable formation and placement of photon emitting centers with narrow line widths, high spectral stability and long coherence times are prerequisites for applications in quantum information science and technology with increasingly complex numbers of qubits and interconnects. Expanding the parameter range accessible for defect engineering with a tunable balance of *in situ* damage formation, ion implantation and annealing supports the search for and optimization of color centers with properties that can be tailored for selected applications, such as emission of (indistinguishable) photons in a telecom band and efficient coupling to a quantum memory for applications in quantum repeaters and quantum networking.

Conventional ion implanters mostly operate with continuous ion beams and current densities in the mA/cm$^2$ range [3,12]. Nanosecond laser pulses and pulsed ion beams with ~100 ns pulse length and peak currents >10 A/cm$^2$ from magnetically insulated diode electrostatic accelerators have been used to process materials, including early studies of simultaneous doping and annealing of silicon [4][13]. Much shorter (sub-ns to ~10 ns) ion pulses can now be generated through the interaction of laser pulses with thin foils at intensities >$10^{18}$ W/cm$^2$ [14]. These short, intense ion pulses from laser-solid interactions are being explored for a broad range of applications [15], including inertial fusion energy [16], nuclear physics [17], (flash) radiation biology [18], and studies of radiation effects in materials and materials analysis [18,19][20][21]. Most of these applications focus on the use of high energy protons and ions (>10 MeV). Here, we report on defect engineering, doping and direct local photon emitter formation in single crystal silicon with the full range of laser-accelerated proton and ion energies, from sub-keV to multi-MeV. We generated ion pulses using the BELLA petawatt laser, a Ti:Sapphire laser (815 nm) [20,22], by focusing laser pulses with pulse energies up to 35 J and pulse lengths down to 35 fs full width half maximum (FWHM) into a ~52 µm focal spot for intensities of 12×$10^{18}$ W/cm$^2$ [23]. Following exposure to a single pulse or to a series of proton and ion pulses, we characterized the resulting changes in the surface structure of single crystal silicon samples and analyzed their local chemical composition, structural and optical properties.

**Results**

**Experimental setup.** Laser pulses were focused onto 13 µm thick Kapton tape targets that had been spooled onto a tape drive so that we could conduct thousands of shots per tape [23] (Figure 1). We conducted laser ion acceleration experiments at a repetition rate of up to 0.2 Hz. Ion spectra were measured *in situ* with a Thomson parabola (TP) type mass spectrometer, complemented by stacks of radiochromic films (RCF) [23,24]. We analyzed samples *ex situ* using secondary ion mass spectrometry (SIMS), nuclear reaction analysis (NRA) and channeling Rutherford Backscattering (ch-RBS) to quantify the fluence and depth distribution of ions that were implanted into the silicon samples as well as the resulting structural changes to the crystal lattice. Ion pulses were dominated by protons and carbon ions from contamination layers on the back surface of the Kapton tapes and from the disintegration of the Kapton foil. High energy proton and carbon ion energy spectra follow the well-known Boltzmann-type distribution from the Target Normal Sheath Acceleration (TNSA) mechanism [23,25], with proton and carbon ion energies up to 8 MeV in our experiments [23]. The silicon samples, positioned 25 mm downstream from the laser target foil, were irradiated using ion pulses with intensities of ~$10^{12}$ ions/cm$^2$/pulse for ions with energies >2 MeV [23] (Figure 1). By analyzing the RCF we found that proton pulses had an energy dependent divergence angle (FWHM) ranging from 8 degrees (7 MeV) to 22 degrees (2 MeV) [18], resulting in an (energy dependent) proton beam spot with a nominal diameter of ~4 to 20 mm at the position of the ion targets 25 mm downstream from the laser targets.

In the top right insert of Figure 1 we show a typical TP mass spectrum with the charge state sequence of ions that is dominated by C$^{4+}$ on the phosphor screen of the multi-channel plate detector. Silicon samples were cut to approximately 1 cm$^2$ chips from n-type float-zone wafers with (111) orientation and a resistance of ~100 Ohm cm. We partially covered the silicon samples with 25 µm thick aluminum foil with a 5 mm diameter aperture. The opening was aligned with the position of the expected peak proton and ion flux. Sample assemblies were mounted together with RCF pieces onto a target wheel that held up to eight samples. We first tuned the laser and laser-ion pulses using the TP and then exposed silicon samples with single pulses, or a series of two, ten, or one hundred pulses of laser-accelerated protons and ions. In the course of repeated ion pulses, we observed evaporation of the aluminum foil around the opening on the silicon samples due to heating by the proton and ion pulses (Supplemental Material). Ion pulses in the TNSA regime have an initial pulse length of picosecond duration. But since the energy distribution of protons and ions is very broad, the TNSA ion pulse de-bunches along the flight path of 25 mm in our setup to about 10 ns (the flight time spread between 8 MeV protons and 0.3 MeV carbon ions).

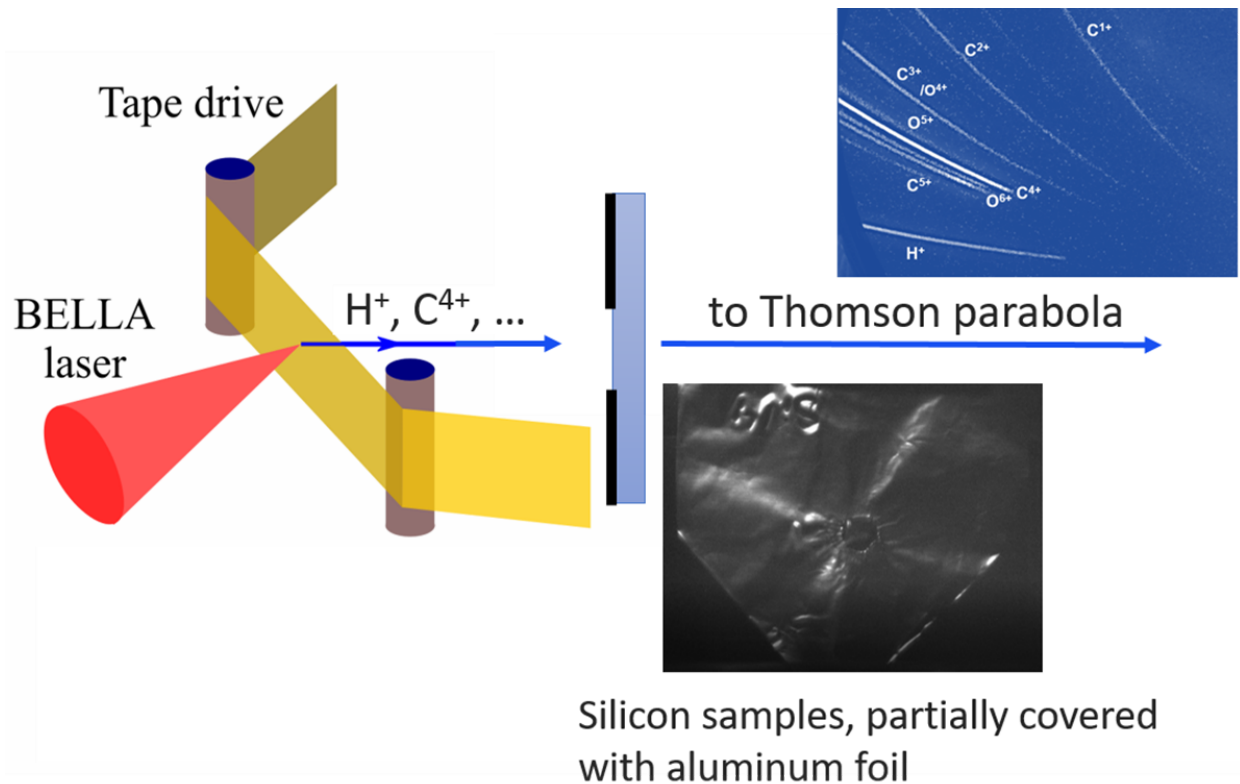

Figure 1: Schematic of the experimental setup (not to scale). (Top right) Example of a Thomson parabola (TP) spectrum dominated by $C^{4+}$ and protons ($H^+$) with energies up to 8 MeV. Following tuning of laser conditions using the TP, we inserted silicon samples for exposure to proton and ion pulses (bottom right). We show a time lapse video of a multi-shot series in the Supplemental Material.

**Surface topography changes and exfoliation.** In Figure 2, we show optical and helium ion microscope (HIM) images (the latter obtained under an angle of 45 degrees) of a silicon (111) sample that had been exposed to one ion pulse. The optical image was taken at a magnification of 6.25x in order to provide an overview of the highly non-uniform surface topography across an area of about 6 $mm^2$. In the center of the image ("D"), we observe a region where layers of silicon were exfoliated. These exfoliated sections are separated by a sharp transition from neighboring areas showing triangular tessellation patterns of cracks that reflect the silicon (111) crystal structure ("C"). Further away from the exfoliated areas, the cracks wane in intensity ("B") and the surface transitions to areas where there is no obvious change in the surface structure ("A"). The field of view in this optical image is 2.45 mm and we show the characteristic features in this single pulse beam spot, which covers part of the 5 mm diameter opening in the aluminum mask. Clearly, the irradiation conditions on the sample were not uniform. The thickness of the exfoliated layers ranged from 4–7 μm, as determined from the step heights between regions "C" and D" measured in the HIM images; see for example the HIM image on the bottom left of Figure 2. The HIM image on the bottom right in Figure 2 shows cracks in the silicon and the onset of an incomplete exfoliation event, with what looks like re-solidified silicon that had been molten and flowed out from under the exfoliating tile. The total area in which we observe exfoliation is about 1 $mm^2$, much smaller than the 20 mm² nominal ion beam spot. The corresponding ion pulse full divergence angle to cover a ~1 mm² spot with peak intensity is only 4.6 degrees, smaller than the earlier result for 8 MeV protons from RCF analysis [18,23]. We note that optical monitoring of damage on a silicon chip can provide rapid *in situ* feedback on proton and ion pulse properties and support tuning of desired laser ion acceleration conditions, complementing TP measurements (which can be affected by the narrow acceptance angle in a TP mass spectrometer). The surface damage results from all ions, and thus also complements the RCF stacks, which are only sensitive to MeV protons. The highly non-uniform damage patterns point to possible structures in the proton and ion pulse from microscopic non-uniformities in the Kapton foil, filamentation effects due the Kapton being insulating, and possible structure in the laser pulse [26,27].

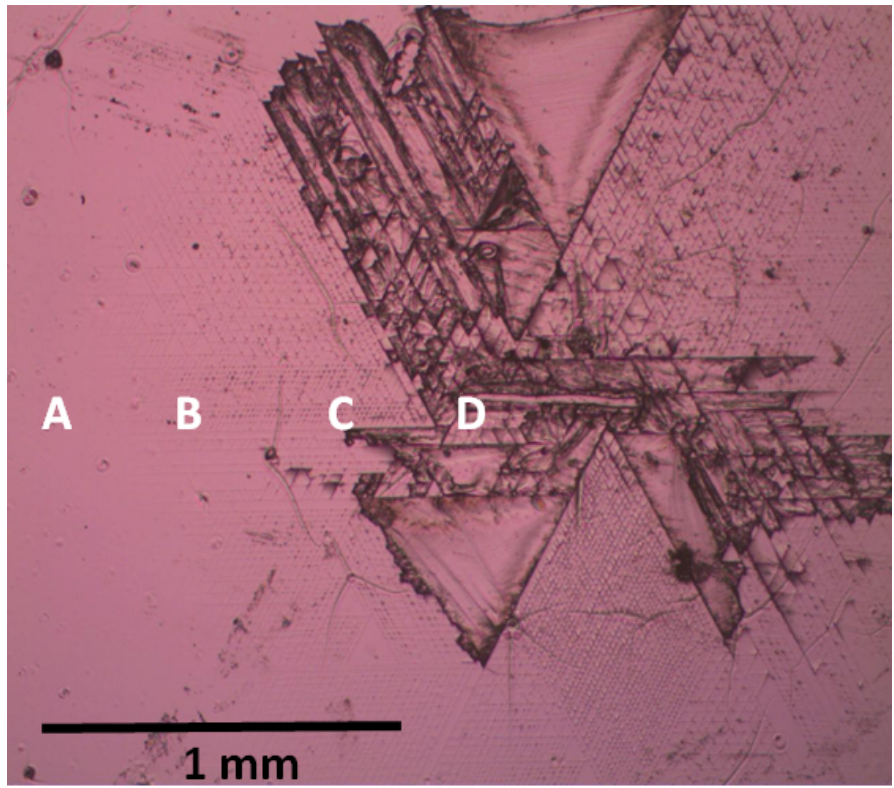

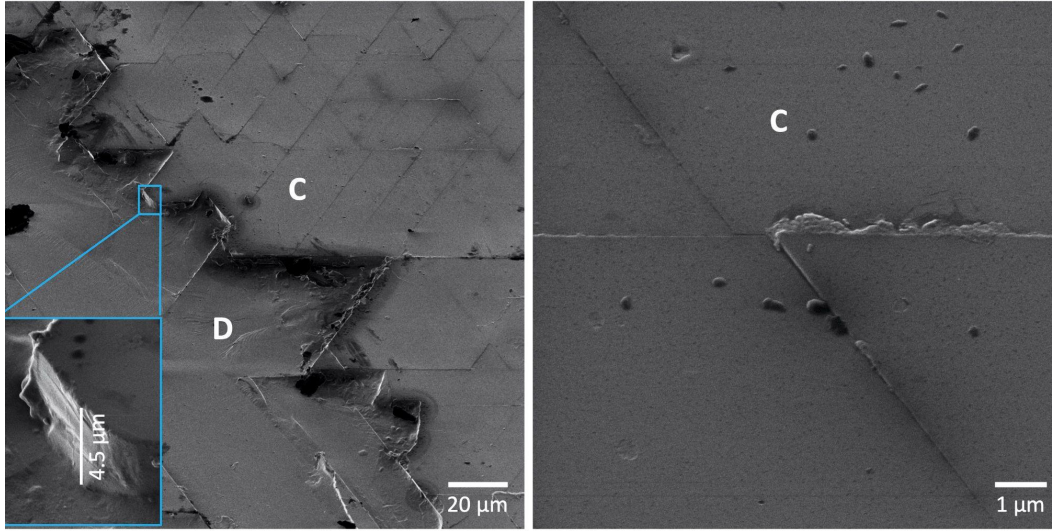

Figure 2: Optical (top) and HIM images (bottom) of a silicon (111) sample that had been exposed to one laser ion pulse. The optical image shows areas with no surface damage ("A") adjacent to areas with tessellation patterns ("B, C") and exfoliation ("D"). (bottom left) Areas with triangular tessellation patterns of cracks and exfoliated tiles of silicon ("C, D"). The insert shows the thickness of an exfoliated layer of 4.5 µm. (bottom right) Area with cracks where a silicon tile started to peel off in response to the local ion flux ("C"). At the top of the exfoliating tile, resolidified silicon that had been molten appears to have emerged from under the tile. Note that the HIM images show the key features of the regions labeled "C, D" but are not of the exact same regions marked in the optical image.

**Ion pulses and local heating of silicon.** Having observed highly non-uniform surface topography changes, we now analyze the local energy deposition from single and repeated ion pulses and estimate the resulting surface temperature profiles in silicon. We characterized the local carbon, oxygen and hydrogen implantation fluences by SIMS in order to complement the TP data (Figure 3). We also note that SIMS can cross calibrate TP spectra. TP based mass spectrometry can capture single shot events with broad ion species and energy distributions, but in order to resolve ion species, a small pinhole aperture is required. For this purpose, we used an opening with a diameter of 0.3 mm about 1 m downstream from the laser target, which limited the field of view of the mass spectrometer. The microscope images above show irradiation with a highly non-uniform ion pulse across the samples and reveal jet-like locally enhanced ion intensities, which can be missed or only partially captured with a TP.

The SIMS sensitivity for carbon in silicon was about $10^{16}$ atoms/cm$^3$ [28]. In order to compare the SIMS results with the TP results, we analyzed a sample that had been exposed to a series of ten ion pulses to accumulate a carbon concentration well above the sensitivity limit of SIMS in the top ~5 µm of the sample. Depth profiles were dominated by carbon, with lower concentrations of oxygen and hydrogen that followed the trend in the local variations of carbon concentrations (Supplemental Material). In Figure 3 (top), we compare SIMS and TP data after normalizing the SIMS data to concentrations per ion pulse. We estimated the depth-energy relation of carbon ions in silicon using Stopping and Range of Ions in Matter (SRIM) code [29]. The slope of the carbon depth profile from SIMS is approximately the same as the slope of the proton and carbon ion intensities as a function of energy above the low energy cutoff of about 3.5 MeV from our TP settings. The SIMS data include all carbon species, while the TP data in this example are for the dominant $C^{4+}$ species.

When calculating the extrapolations shown in the data in Figure 3 we assume that the dominant contribution to the proton and ion flux is from TNSA and extrapolate the exponential distribution to zero ion energy. The resulting local, peak energy deposition from energetic protons and ions is about 2 J/cm$^2$. But other mechanisms can also contribute to the low energy ion flux, including plasma expansion [30] and disintegration of the Kapton foil. In samples that had been exposed to single ion pulses we observe carbon concentrations up to 10$^{20}$ atoms/cm$^3$ with a peak depth of 200 to 500 nm. If these carbon atoms were implanted with average ion energies of 50 to 200 keV (corresponding to an implantation depth of 150 to 500 nm) then the corresponding energy deposition for a corresponding fluence of ~10$^{15}$ ions/cm$^2$ would locally each ~30 J/cm$^2$. Analytical calculations and finite element analysis of the corresponding temperature rise show heating to temperatures well above 10$^4$ K for intensities of 30 J/cm$^2$, which seems unrealistically high given the observed surface structures. We provide details on the temperature profile calculations in the Supplemental Material. We observe these high carbon concentrations with locally varying peak concentrations and depth distributions across the samples. The series of SIMS depth profiles in Figure 4 corresponds to the respective areas labeled in the optical image in Figure 2. Region "A" was near the edge of the 5 mm diameter nominal beam spot. The SIMS profile shows a very high concentration of carbon near the surface. The areal density is 3–8×10$^{15}$ C/cm$^2$, over ten times higher than from a control sample. This high fluence of near surface carbon stems from the deposition and implantation of very low energy carbon ions that are associated with the plasma expansion of the Kapton foil following the intense laser pulse. In Figure 3 (bottom) we show calculated temperature profiles as a function of time based on the ion energy deposition extracted from the SIMS and TP spectra and using known values for the thermal conductivity of silicon (Supplemental Material). The high peak temperature of 3750 K can be considered a lower limit based on the extrapolation of the ion spectra. We find that the peak temperature drops below the melting point of silicon after about 1 µs following the ~10 ns long ion pulse. Given the thermal conductivity and heat capacitance of silicon, the peak temperature reached at a depth of 6 µm roughly equals the melting temperature. This depth coincides with the thickness of the exfoliated layers (Figure 2). We observe exfoliation on samples that had been exposed to one single ion pulse, as well as for a series of ion pulses. The thickness of exfoliated layers is similar to that reported earlier for a series of over ten 100 ns long ion pulses from an electrostatic ion accelerator and energy deposition of about 1 J/cm$^2$ [31] where the peak silicon temperature had been evaluated to be 1600 K, about two times lower than in our estimate.

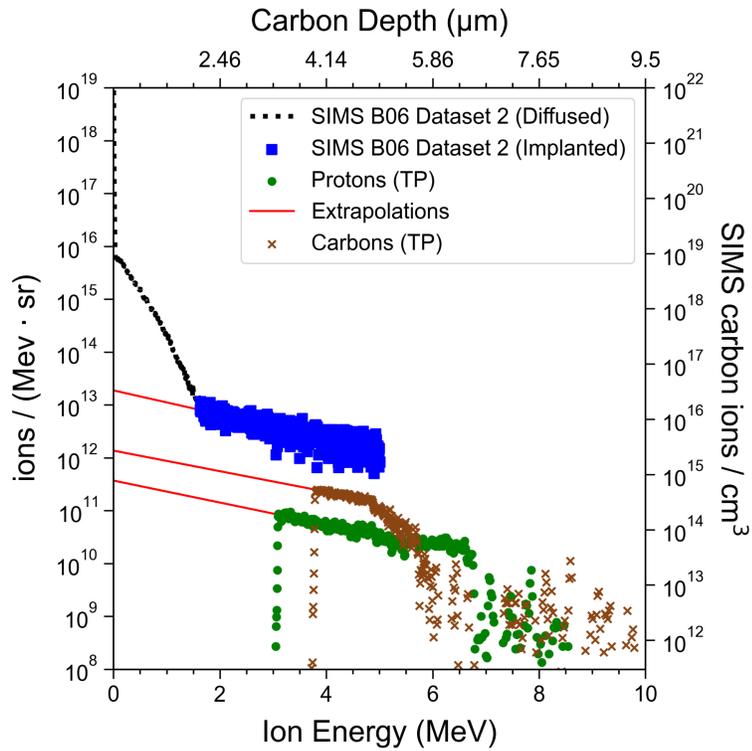

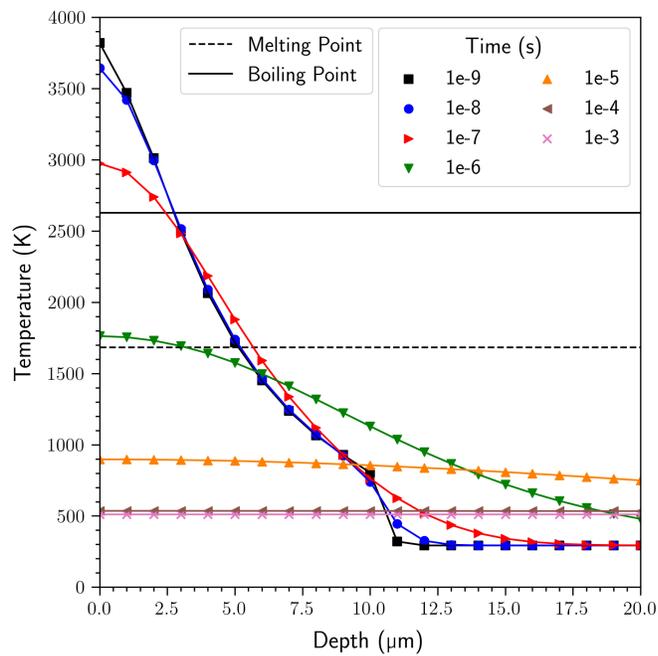

Figure 3: Top: Examples of TP data for protons and $C^{4+}$ ions, together with a SIMS depth profile for all carbon ion species. The red lines show extrapolations of the TP and SIMS data to low energies. Bottom: Sample temperature as a function of depth for a series of time steps following a laser ion pulse. Details of the energy deposition to temperature profile calculations are in the Supplemental Materials.

The superposition of MeV protons and carbon ions from TNSA and a pulse of much lower energy ions from plasma expansion poses a challenge in understanding the local energy deposition and temperature evolution.  To elucidate the relative contributions of TNSA ions and ions from plasma expansion, we analyzed samples that had been exposed to 10x and 100x ion pulses by Nuclear Reaction Analysis (NRA) [32].  NRA complements SIMS measurements of local carbon concentrations at a series of positions across a sample that had received one pulse.  Details on NRA are in the Supplemental Materials.

Figure 4 (top) shows the carbon concentrations of samples from 10x and 100x ion pulses as a function of depth for a series of analysis spots across the sample.  NRA is based on inducing nuclear reactions between MeV projectile ions and host atoms in the sample.  NRA is thus insensitive to chemical effects, or so called matrix effects, that can affect SIMS depth profiles in the presence of very high impurity concentrations and in the presence of surface roughness, which are both present in our samples.  But the sensitivity, lateral (~0.1 mm) and depth resolutions (a few nm) of SIMS far exceed those of NRA.  NRA shows peak carbon concentrations equal to the density of silicon atoms, i. e. $5\times10^{22}$ cm$^{-3}$, in the top 200 nm.  We observe carbon at a depth greater than 200 nm only for the central area of the sample ("spot 5" for the 100x sample).  In this central area, carbon is found at a depth as deep as $7.5 \pm 1.5$ μm at a concentration of $8.3\times10^{18}$ C/cm$^3$/pulse (where we have normalized the concentrations to the number of ion pulses).  The line scan of NRA profiles across the 10x and 100x pulse samples shows a profile of carbon areal densities peaking at $1.7–2.5\times10^{16}$ C/cm$^2$/pulse.  When we plot the areal density of carbon at a depth larger than 200 nm for the 100x pulse sample we see that it is localized near the center of the sample (spot 5).  A fraction of these carbon atoms stem from high energy ions that were implanted.  But most of the carbon is very close to the surface.  In the center of the samples, where we also observe the most intense damage, we see carbon to a depth of over 7 microns.  We argue that most of the carbon comes from plasma expansion and the disintegration of the Kapton foil.  MeV ions from TNSA reach the silicon sample after a few ns and locally melt the silicon.  Low energy carbon ions with energies as low as ~100 eV reach the silicon sample after a flight time of ~1 μs, the time when the temperature drops below the melting point in our estimate.  Once low energy carbon ions impinge on the hot and molten silicon surface, they can rapidly diffuse and redistribute, until the silicon cools and (partially) recrystallizes after >1 μs.

We complement NRA scans (with analysis spots of 1 mm diameter, along a 10 mm long line scan across the sample) with SIMS depth profiles (with analysis spots of 0.1 mm diameter, line scan over ~1 mm on the sample) in Figure 4 (bottom).  The SIMS spectra of a sample that had received one ion pulse (Figure 2) show a progressively increasing depth of carbon atoms with a slowly varying integrated areal density (insert, Figure 4).  A control sample that had not been exposed to ion pulses showed a near surface profile of carbon with an areal density of $3.5\times10^{14}$ C/cm$^2$, much lower than the observed carbon areal densities across exposed samples.  The deepest carbon profiles are observed closest to the highly damaged area.  Inside this exfoliated area, SIMS depth profiles became unreliable due to the highly textured surface.  This trend in the SIMS profiles is consistent with the NRA result and supports the interpretation of local carbon drive-in and diffusion, with a smaller contribution of high energy carbon ions. Following 1x, 10x and 100x ion pulses, carbon atoms accumulate near the surface of samples and are driven progressively deeper into the silicon.

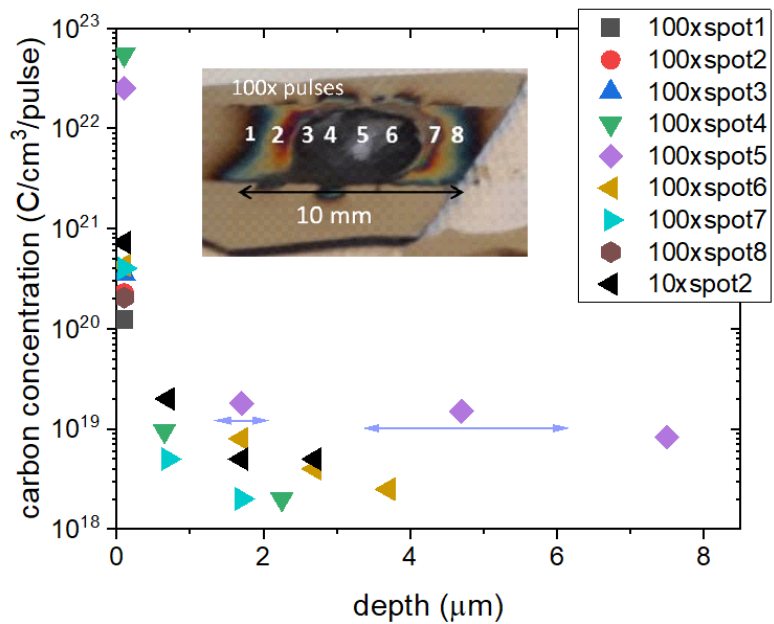

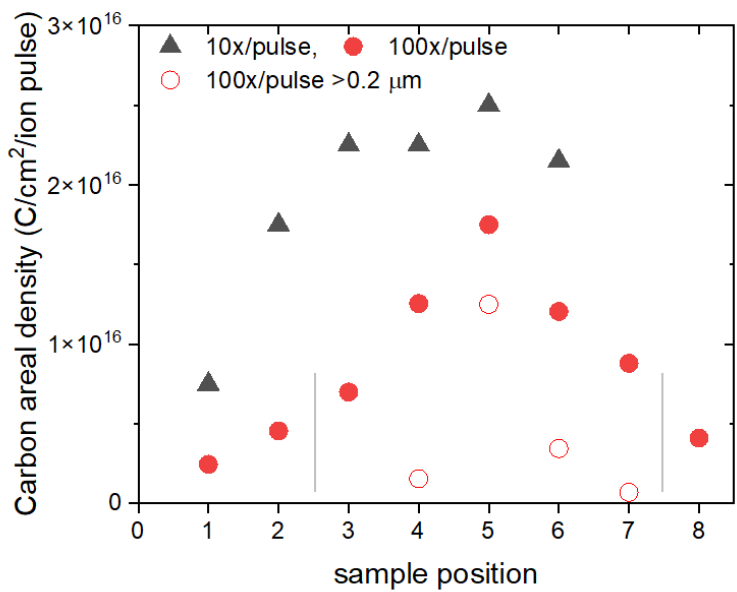

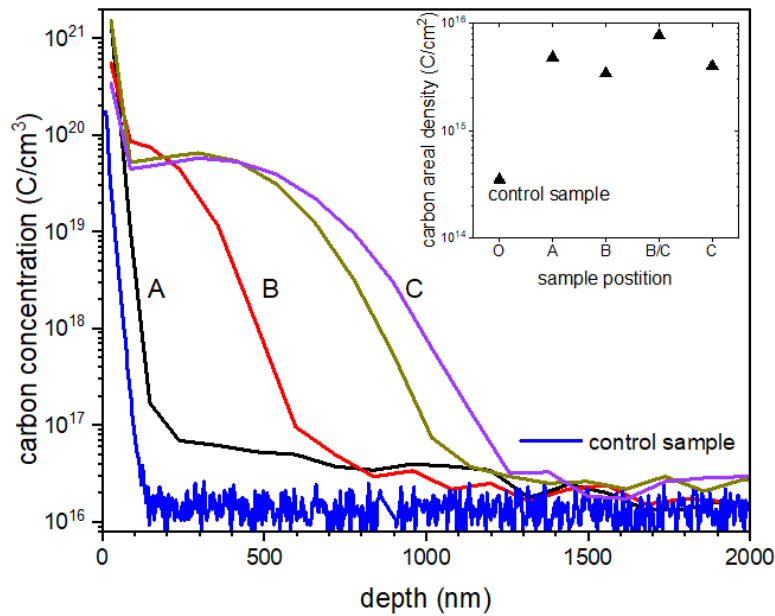

Figure 4: Top: carbon concentration as a function of depth for the NRA analysis spots on the 100x pulse sample and for the central beam area on the 10x pulse sample ("spot 2", Supplemental Material). The depth intervals of 1 to 3 µm used for signal integration are indicated by the arrows. Mid: Carbon area density near the surface (top 200 nm) for the 10x and 100x pulse samples for the series of NRA analysis spots across the laser ion beam spot in a 10 mm line scan. For the 100x pulse sample we also show the carbon areal density integrated for depth >200 nm. The concentration and areal density values are normalized to the number of pulses for comparison. The gray lines indicate the initial 5 mm wide opening in the aluminum foil. Bottom: SIMS depth profiles of carbon in a silicon sample that had been exposed to one laser ion pulse for a series of positions across 1 mm of a sample as indexed in Figure 2. The insert shows the carbon areal densities for the series of positions.

Low energy carbon atoms are thus driven into the silicon sample due to the high local energy deposition and heating from the high energy ion pulse(s). This is analogous to laser driven atom in-diffusion into silicon with laser pulses of similar duration (few ns) and intensity (few J/cm$^2$) [33] and with surface adsorbate layers as the source of the atoms that are diffused into the material [34]. But a major difference is that the doping atoms, here carbon, originate from the plasma expansion of the laser foil target (here Kapton). A choice of a different material for the laser foil target will enable high flux, high fluence doping with other selected species (e. g. boron). We thus observe a new high flux surface doping process, "laser-ion doping", that is complementary to gas immersion laser doping. In laser-ion doping, we observe peak surface dopant concentrations well in excess of 1 atomic % from single ion pulses, compared to the tens of pulses reported for gas immersion laser doping required to achieve similarly high concentrations. The characterization of electrical properties of materials with very high dopant concentrations from laser-ion doping is subject of ongoing studies.

We complemented microscopies of surface topography changes and the compositional analysis of the samples by SIMS and NRA with channeling Rutherford Backscattering (ch-RBS) measurements (Supplemental Material) to track the structural evolution of silicon crystal samples for a series of 1, 2, 10 and 100 ion pulses. Ch-RBS is sensitive to the areal density of displaced atoms [35]. Ch-RBS showed that the density of displaced silicon atoms increased with increasing number of ion pulses. But point defect concentrations stayed well below levels that would signify the presence of amorphous silicon layers from the accumulation of radiation damage, consistent with damage annealing in the high thermal budget from ion energy deposition, and consistent with the formation of dislocation loops in the highly carbon doped silicon samples.

**Color center spectra.** Following exposure to a single pulse or to a series of ion pulses, we conducted low temperature photoluminescence (PL) measurements to characterize the resulting color center spectra. In Figure 5, we show a 5 mm wide line scan of PL spectra taken at 3 K across a sample that had been exposed to two laser-ion pulses. We used continuous wave excitation at 532 nm for the PL measurements. Comparing the results from several samples that had been exposed to the same number of ion pulses we found that the results were reproducible, though the particular local irradiation conditions, surface changes and resulting color center distributions varied. We found that the PL spectra were dominated by emission lines at 1217 nm and 1280 nm, corresponding to the well-known W and G-centers in silicon, respectively [6] [7] [8].

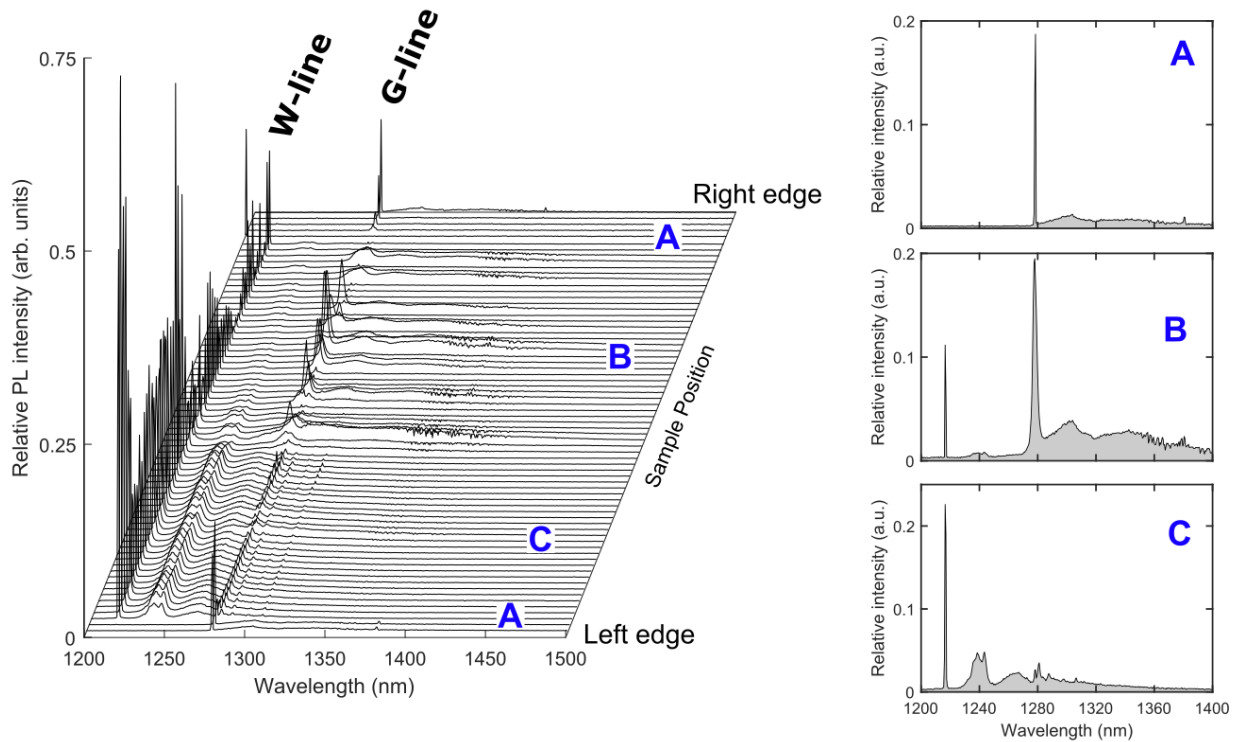

Figure 5: Left: Line scan of low temperature PL spectra taken across a silicon (111) sample that had been exposed to two laser-ion pulses. Right: Spectra form the left and right edge of the sample where we observe only G-centers, corresponding to the area labeled "A" in the optical microscope image shown in Figure 2. Areas with increased ion

flux show both W and G-centers, corresponding to area "B" in Figure 2. Regions with higher local ion flux show mostly W-centers ("C" in Figure 2).

We see an intense signal from G-centers at 1280 nm from a region on the sample that had been masked by the 25 μm thick aluminum foil during the irradiation. The aluminum foil stopped all carbon ions and also protons with energies up to about 1.4 MeV. The spectrum of higher energy protons is downshifted by energy loss in the foil. This results also in a shortening of the proton pulse to sub-ns duration, because the low energy tail of protons is stopped in the aluminum foil. Energetic protons form silicon interstitials in close impact parameter collisions and these can diffuse towards the surface where they can combine with substitutional carbon pairs, forming G-centers [36]. We do not know the exact spectrum of protons because in our measurements the TP has a low energy cutoff of about 3 MeV. We do know the concentration of carbon near the surface from SIMS analysis of control samples and samples that had been covered with aluminum foil (Figure 4). Formation of G-centers from irradiation of silicon that contains carbon is well known (Davies 1989). Direct local formation of G-centers with carbon that was already in the lattice and silicon interstitials formed by energetic protons represents an elegant way to locally add G-centers into silicon and potentially into integrated silicon devices without any subsequent thermal annealing. The high flux of protons, estimated to be $10^{11}$ protons/cm$^2$/ns from the TP data and given the divergence angle of the proton pulse, can aid G-center formation if the mobility of silicon interstitials is increased due to the transient local excitation and heating of the matrix (Schenkel et al. 2022). The ability to tune the proton flux from a compact laser accelerator now extends the parameter space for local G-center formation.

In areas that had been exposed to both proton and carbon ions we observe spectra with G as well as W-centers (Figure 5, spectrum "B" on the right). Comparing results from several single and two pulse samples we find that W and G-centers are formed in areas of intermediate ion flux and local heating, areas "B" in Figure 2. In regions with higher local energy deposition, only W-centers are observed (area "C"). Here, we find that the line width of the G-centers increases significantly in areas of higher surface damage and higher carbon concentrations from higher ion flux conditions, while the width of the W-center peak does not increase. The transition from areas with both G and W-centers to those with only W-centers also maps to the relative temperature stability of these centers, with G-centers being significantly more likely to dissolve at temperatures as low as 470 K, while W-centers have been reported to be stable up to at least 600 K[6–8]. These temperatures from earlier studies appear very low considering the much higher peak temperatures we estimate from our ion pulse conditions. But this points to the important difference in thermal budgets applied over an extended period (>1 s) in conventional thermal annealing in contrast to the condition of very rapid heating (~3700 K in 10 ns, or >$10^{11}$ K/s) and cooling (~3x$10^9$ K/s) where specific defect configurations can be formed and stabilized during rapid quenching before they dissolve. PL spectra are shown in the Supplemental Material together with SIMS depth profiles that had been taken prior to PL at the same locations on the sample.

**Density Functional theory (DFT) of G and W-centers vs. local disorder.** Having observed local color center formation for a series of irradiation conditions we conducted DFT calculations to understand the sensitivity of W and G-centers to local disorder. To model the electronic properties of these defects we use the G-center Type-B [29,37 38 39 40] and W-center Type-V [40,41 42 42,43] structures. To determine the sensitivity of the defect energy levels to perturbations in the local environment, these structures were first relaxed, and then used as a starting point for quantum molecular dynamics (QMD) simulations. We let the structures evolve for 3 to 5 ps after thermalization and collect the defect energy levels at each time step. These energies are computed using the PBE functional (see Methods), which is known to underestimate the magnitude of the insulating gap. Nevertheless, the fluctuations around the average value are expected to reproduce the correct trends for these defect centers.

We use QMD simulations to understand the sensitivity of G-center and W-center defects to local disorder. While the structural fluctuations arise from thermal effects in our simulations, they can also be used to probe the dependence of optical properties on inhomogeneous broadening effects that arise due to static disorder. This is because local disorder, induced by other distant defects in the dilute limit, can be understood as low-energy distortions which are well-represented by thermal fluctuations of the QMD trajectories. The QMD simulation results in Figure 6 show two apparent trends. Firstly at each temperature, the G-center distributions are broader than for the W-center, consistent with the observed narrower linewidth of the W-center emission spectra reported above. Secondly, Figure 6 shows that the effect of increasing temperature, and hence, increasing disorder, is more pronounced for the G-center than the W-center. At lower temperatures the slope of the spread is steeper for the G-center than for the W-center. It should be noted that this estimate of linewidth broadening is limited by the finite size of the simulation cell. Periodic boundary conditions place defect centers in close proximity to their images in adjacent unit cells, leading to possible interactions, while the finite size sets a limit on the allowed phonon-frequencies. These effects lead to a systematic overestimation of the broadening for both types of defect centers, but the trends and relative magnitudes can still be compared. The main conclusion that can be drawn is that in order to obtain G-centers with narrow linewidths, low disorder will be essential, which informs strategies for process optimization. On the other hand, the W-center is more robust to disorder, which is a promising property given that the creation of a W-center requires the formation of three silicon interstitials and is often associated with high damage events and processes.

The effect of excess interstitial carbon on the electronic structure of the defects was also analyzed. Neutral carbon defects were inserted at each of the 106 possible tetrahedral interstitial sites [36] within the 3 × 3 × 3 silicon supercells hosting the W and G-center defects. Excess carbon interstitials have a similar effect on the two defect structures. In the Supplemental Material, we include scatter plots showing the defect gaps and energies for varying distance between carbon interstitial and defect center. Here the trend is more subtle. While, unsurprisingly, both centers have significant variation in their defect properties when carbon interstitials are in close proximity, in the real samples, the carbon concentration of order 1 atomic % are reached only when in very close proximity to the sample surface (i. e. the top ~200 nm, while the PL probes a depth of ~1 μm). When considering the configurations where the carbon interstitial is farther away (10–12 Å), the properties of the G-center experience much greater variation than those of the W-center. The relative stability of the G-center and W-center to perturbations can be understood from a structural standpoint. Currently, the structure of the G-center is believed to be the B-type configuration, which can be seen as two substitutional carbons and a twofold coordinated interstitial silicon between them, forming a $C_s$-$Si_i$-$C_s$ dumbbell structure. When this structure was first proposed [37], it was already noted that this dumbbell structure has 3 possible orientations around the substitutional carbons, explaining the $C_{3v}$ rotational symmetry observed in the EPR spectrum. It was also suggested that this center could transform to a related A-type structure 0.02 eV higher in energy through a bond switching process which would generate threefold coordinated $Si_s$ and $C_s$ [37]. More recently, it has been shown that the low coordination of the B-type structure allows it to rotate about the [111] axis through several local minima [9,37]. Conversely, each silicon atom comprising the W-center structure is covalently bonded to four nearest neighbors in a tetrahedral arrangement, which is much less susceptible to structural distortions.

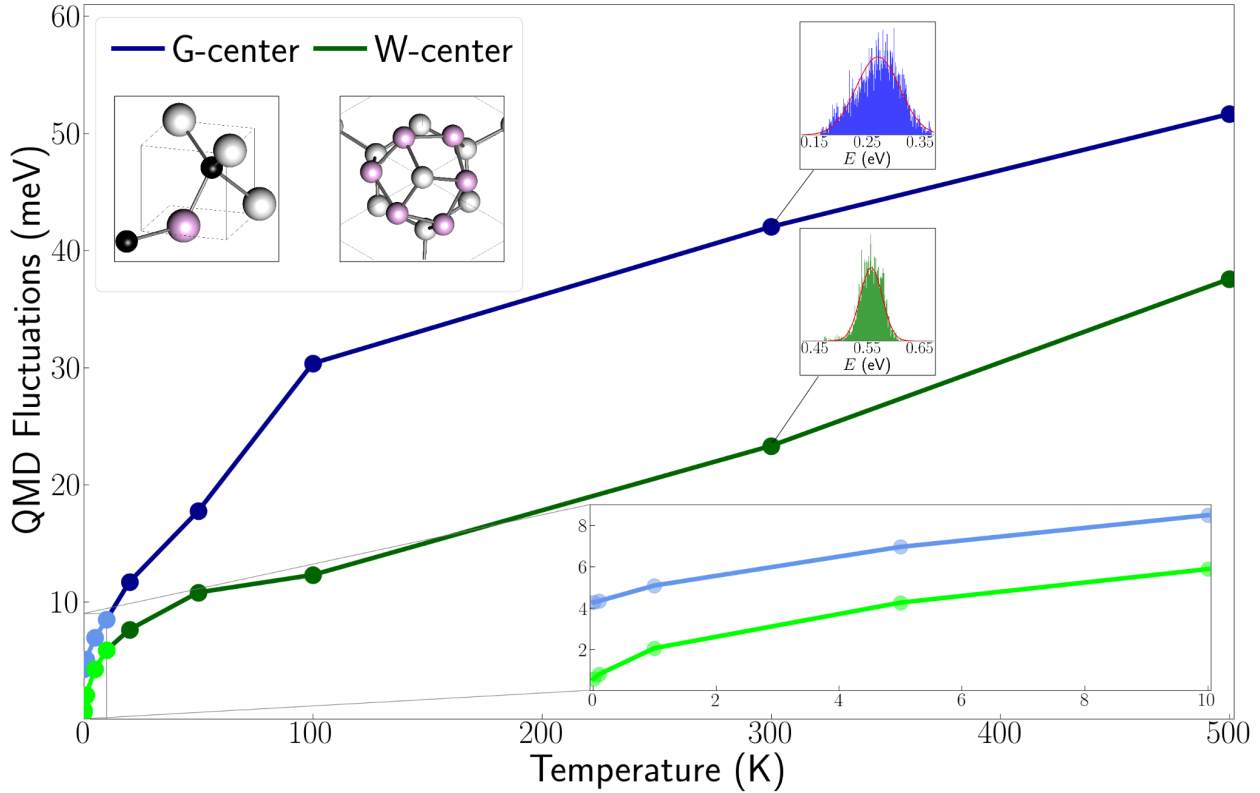

Figure. 6: QMD fluctuations for W and G-centers vs. simulation temperature, which represents local disorder. The G-center shows a higher sensitivity to disorder than the W-center, in agreement with experimental results. Insets (top left) show G and W-center structures with C-atoms in black, Si-atoms in gray, and defect (interstitial) Si-atoms in pink. Each point in the plot represents ~5000 configurations obtained through QMD. Inset (bottom right) shows a magnified region of the plot for 0–10 K. Sample histograms (top right) of this data are shown at 300 K for both defect centers.

**Discussion**

Analysis of silicon samples exposed to single pulses or a series of pulses of laser accelerated ions shows two distinct contributions to the total ion fluence. The first contribution is from MeV ions that form an intense, short (1 to 10 ns) pulse from laser acceleration through TNSA with a typical fluence of ~$10^{12}$ cm$^{-2}$ and local energy deposition up to ~2 J/cm$^2$ in our experimental setup. A second pulse of very low energy ions and particles, with kinetic energies <1 keV, is formed through plasma expansion and the disintegration of the laser-foil target (here Kapton). We find that the fluence of these low energy ions and particles is ~$10^{16}$ cm$^{-2}$, about four orders of magnitude higher than that for protons and ions from TNSA. The low energy ions trail the MeV ions, and can arrive at the silicon samples when it is still hot or locally molten. We estimate that 100 eV ions can reach the silicon surface within 1 μs in our setup, fast enough to then quickly diffuse into molten silicon. This high flux surface doping process is analogous to gas immersion laser doping [34][33], but now with spatio-temporal alignment of a high energy ion pulse followed by a pulse of low energy dopant ions and particles in a laser accelerator. This finding opens a new direction for exploration of high flux surface doping and synthesis of materials [34,44]. We observe carbon concentrations in excess of several atomic % per ion pulse near the surface of silicon and this suggests that very highly doped materials phases might become accessible with this approach. Laser-foil targets, including boron or other dopants, can be prepared *in situ* to reduce the effect of common surface contaminants for synthesis of materials with tailored properties (e. g. superconductivity in covalent semiconductors with high boron concentrations) [27,34, 44].

We observe highly non-uniform surface topography changes across the 5 mm nominal beam spot. The regions where silicon is exfoliated are structured and do not represent simple Gaussian ion pulse shapes. High flux regions showing exfoliation cover about 1 mm$^2$, indicating full divergence angles in the high energy ion pulses of ~4.5 degrees. *In situ* optical microscopy at low magnification (~5 to 10x) can be used in future studies to characterize the local ion intensity using damage on silicon targets as an observable in support of rapid tuning of laser parameters and optimization of the experimental setup.

Exfoliation of silicon (100) following exposure to ~50 ion pulses from an electrostatic accelerator and from exposure to high flux low energy plasma flows have been reported earlier [31, 45]. Here, we find local exfoliation from single ion pulses. In laser-ion acceleration using TNSA, the carbon and proton ions de-bunch to about 10 ns in our setup with an ion flight distance of 25 mm. This pulse length is shorter than the pulses used in earlier studies[31] using a conventional ion acceleration mechanism. The depth of the exfoliated silicon layers is the same for both types of pulsed ion beams and plasmas, reflecting the thermal properties and response of silicon to rapid surface heating in short ion pulses. The sharp transition from exfoliated areas to areas that exhibit cracks in tessellation patterns point to a phase transition across the melting point for the onset of exfoliation. Our analysis of the ion energy deposition and resulting temperature increase shows that the sample cools below the melting point within about 1 µs. But measurements of the energy dependent local ion flux contributions from TNSA and plasma expansion have to be improved to refine these estimates.

We observe bright emission from G-centers that are formed locally by proton pulses and with carbon that is already present in the silicon lattice [46]. A tunable flux of laser-accelerated protons could be directed to selected regions using e. g. a plasma lens [18,33] to add G-centers into completed silicon devices (most of which contain some carbon, especially near the surface) with a compact laser accelerator. This local addition of color centers with MeV protons also occurs naturally with protons and other high energy cosmic radiation, and can be considered as a hardware error mode for future quantum devices that include color centers in silicon. Quantifying the radiation hardness of emerging quantum devices is becoming increasingly important as qubit integration advances and can also be addressed with compact laser accelerators.[47,48]

Color centers form directly in response to local excitation where the balance of damage formation and annealing leads to the formation of W and G-centers with varying relative intensities. Rapid local heating and quenching on the 10 ns and 1 µs time scale, respectively, can aid formation and stabilization of color centers that would dissolve during conventional thermal annealing steps of much longer duration. DFT calculations of color center emissions as a function of local disorder are in agreement with the experimental observation of stronger linewidth broadening of G-centers vs. W-center ensembles in high ion flux regions. We note that understanding this broadening underpins the application of these color centers for local measurements of local damage and stress fields in silicon, as well as their potential application in high radiation environments.

Our results inform novel strategies for defect engineering in semiconductors, including the formation of highly doped surface layers and the integration of color centers. With increased control of local ion flux conditions, intense, pulsed ion beams from (compact) laser accelerators extend the parameter range accessible for materials synthesis, and the exploration of novel materials phases and color centers that can be formed and stabilized under conditions of rapid local excitation, high flux doping and quenching.

**Methods**

**Petawatt laser system**. The experiment was performed using the BELLA PW laser facility at LBNL. The BELLA PW laser is the world's first 1 Hz repetition rate 1 PW Ti:Sapphire laser system based on double-chirped pulse amplification architecture, where a cross-polarized wave (XPW) contrast enhancement system is installed in between two CPA stages, delivering pulses with a duration down to ~35 fs FWHM at 815 nm central wavelength. A 13.5 m focal length off-axis parabolic mirror is used to focus the laser pulses with around 35 J energy to a measured spot size of 52 μm FWHM diameter, yielding a peak intensity of $12 \times 10^{18}$ W/cm$^2$. For this experiment the laser was operated at 45 fs pulse length, optimized for maximum proton energy. Although 1 Hz operation would be possible, the experiments were performed with a repetition rate of 0.2 Hz.

**Tape-drive target**[49]. Kapton tape with a thickness of 13 μm was irradiated under a 45 degree angle of incidence. In our target assembly, the Kapton tape is spooled into a feedback-controlled tape drive system, and moved by two DC-motors, providing a fresh wrinkle-free target surface with a position repeatability < 10 μm. The tape drive is capable of operating at a repetition rate of 1 Hz.

**Helium Ion Microscopy**. A Zeiss ORION NanoFab instrument was used to image silicon samples using 25 keV helium ions with a beam current of 1.35 pA and a nominal beam spot size of 0.5 nm. Images were acquired using an Everhart-Thornley detector to collect secondary electrons for a sample tilt angle of 45 degrees.

**Secondary Ion Mass Spectrometry**. SIMS analysis was conducted by colleagues at EAG (www.EAG.com).

**Rutherford backscattering analysis.** We conducted RBS measurements with channeling analysis using 2 MeV helium ions and 1 mm beam spots.

**Nuclear Reaction Analysis.** We conducted NRA measurements on silicon samples with 1.4 MeV deuterium ion beams and using the (d, p) reactions, also with 1 mm beam spots.

**Heat analysis.** We conducted the analysis of sample heating for a series of ion pulse conditions using the SRIM code (www.srim.org) to extract energy deposition profiles and applying finite element analysis methods.

**Photoluminescence.** The PL spectra were recorded at 3K using a scanning confocal microscope optimized for near-infrared spectroscopy. Optical excitation was performed with a 532 nm continuous wave laser focused onto the sample through a high-numerical-aperture microscope objective (NA=0.85). The excitation power, measured at the entrance of the objective lens, was fixed at 1mW. The PL signal was collected by the same objective and directed to a spectrometer coupled to an InGaAs Camera (900–1620 nm at -80C).

**Density Functional Theory.** Numerical simulations of silicon defects were performed using the Vienna ab initio Simulation Package (VASP) [50] [51] [51,52] [53]. Initial defect structures were relaxed using the HSE06 functional [40] [54] using an energy cutoff of 400 eV and converged within an energy tolerance of $10^{-8}$ eV and forces of 0.01 eV/ Å at the Gamma point. Quantum molecular dynamics (QMD) simulations used a finite temperature canonical ensemble with an Andersen thermostat [55]. Modified defect structures are relaxed using PBE functionals [56] and the same tolerance parameters used for the original structures.

**Data availability.**

Data will be made available for reasonable requests.

**Acknowledgments**

This work was supported by the Office of Science, Office of Fusion Energy Sciences, of the U.S. Department of Energy, under Contract No. DE-AC02-05CH11231. Experiments at the BELLA Center were enabled through facilities developed by HEP and LaserNetUS. We (TS, JGL) gratefully acknowledge support by the coordinated research project "F11020" of the International Atomic Energy Agency (IAEA). LZT and JS were supported by the Molecular Foundry, a DOE Office of Science User Facility supported by the Office of Science of the U.S. Department of Energy under Contract No. DE-AC02-05CH11231. This research used resources of the National Energy Research Scientific Computing Center, a DOE Office of Science User Facility supported by the Office of Science of the U.S. Department of Energy under Contract No. DE-AC02-05CH11231. Helium Ion Microscopy was performed at the Biomolecular Nanotechnology Center, a core facility of the California Institute for Quantitative Biosciences at UC Berkeley. We would like to express our special thanks to the technical support team at the BELLA Center, Arturo Magana, Joe Riley, Zac Eisentraut, Mark Kirkpatrick, Tyler Sipla, Jonathan Bradford, Nathan Ybarrolaza and Greg Manino. We gratefully acknowledge support for RBS analysis at the Laboratory for Surface Modification, Rutgers University.

**Author contributions**

WR conducted low temperature PL measurements; AJA, PAS, AP and BW conducted ion pulse and energy deposition analysis; FIA conducted HIM measurements; IP conducted SEM measurements; TS conducted optical microscopy; LOH, KN, SS, JHB, QJ, AG, LG and TS conducted laser ion acceleration experiments; SB and CS provided theoretical support on laser-ion acceleration; JGL and JFF conducted NRA measurements; HH and LF conducted ch-RBS measurements; LZT, JS and VI conducted DFT calculations; RW, GB and JM provided early support on color center data acquisition and interpretation; BK, CG, CS, EE and TS provided supervisory support. All authors contributed to data analysis and the writing of the manuscript.

**Competing interests:**

The authors have no competing interests.

**Supplemental information - Defect engineering of silicon with ion pulses from laser acceleration**

Walid Redjem[1], Ariel J. Amsellem[2], Frances I. Allen[3, 4], Gabriele Benndorf[5], Jianhui Bin[2,6], Stepan Bulanov[2], Eric Esarey[2], Leonard C. Feldman[7], Javier Ferrer Fernandez[8], Javier Garcia Lopez[8], Laura Geulig[2, 10], Cameron R. Geddes[2], Hussein Hijazi[7], Qing Ji[2], Vsevolod Ivanov[2,9], Boubacar Kante[1,2], Anthony Gonsalves[2], Jan Meijer[5], Kei Nakamura[2], Arun Persaud[2], Ian Pong[2], Lieselotte Obst-Huebl[2], Peter A. Seidl[2], Jacopo Simoni[9], Carl Schroeder[2,11], Sven Steinke[2,12], Liang Z. Tan, Ralf Wunderlich[5], Brian Wynne[2], and Thomas Schenkel[2, 11, ]*

[1]Department of Electrical Engineering and Computer Science, University of California, Berkeley, CA 94720, USA

[2]Accelerator Technology and Applied Physics Division, Lawrence Berkeley National Laboratory, Berkeley, CA 94720, USA

[3]Department of Materials Science and Engineering, University of California, Berkeley, CA 94720, USA

[4]California Institute for Quantitative Biosciences, University of California, Berkeley, CA 94720, USA

[5]Felix-Bloch Institute for Solid State Physics, Department: Applied Quantum Systems, Linnéstr. 5, D-04103 Leipzig, Germany

[6]present address: State Key Laboratory of High Field Laser Physics and CAS Center for Excellence in Ultra-Intense Laser Science, Shanghai Institute of Optics and Fine Mechanics, Chinese Academy of Sciences, Shanghai 201800, China

[7]Rutgers University, Department of Physics & Astronomy, 136 Frelinghuysen Rd, Piscataway, NJ 08854, USA

[8]University of Seville, Seville, Facultad de Física, Centro Nacional de Aceleradores, 41092 Sevilla, Spain

[9]Molecular Foundry, Lawrence Berkeley National Laboratory, Berkeley, CA 94720, USA

[10]present address: Fakultät für Physik, Ludwig-Maximilians-Universität München, Am Coulombwall 1, 85748 Garching, Germany

[11]Department of Nuclear Engineering, University of California, Berkeley, CA 94720, USA

[12]Present address: Marvel Fusion GmbH, Blumenstrasse 28, D-80331 Munich, Germany

*corresponding author: T_Schenkel@LBL.gov

1. **Time lapse movie showing evaporation of the aluminum foil mask during 100 shots.**

2. **Details on energy deposition and heat calculations**

3. **Details on NRA**

4. **Details on ch-RBS**

5. PL and SIMS data correlation for the PL data shown in Figure 4

6. Supplemental material on DFT calculations of G and W-centers in silicon

1. Time lapse movie showing evaporation of the aluminum foil mask during 100 shots

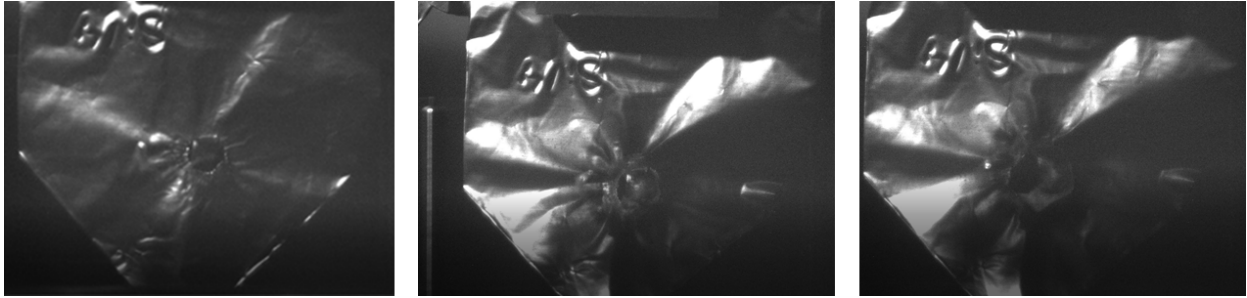

Movie S1: The movie (posted online) is compiled from video camera images taken during a series of 100 shots at the BELLA PW laser where laser-accelerated protons and ions irradiated a silicon sample that was covered with aluminum foil (Figure 1, setup in main text). The real time repetition rate was 0.2 Hz. The film shows the gradual evaporation of the aluminum foil around the central aperture with an initial diameter of 5 mm. Link to the movie: https://youtu.be/1Lxy5yUgQQY

2. Details on energy deposition and heat calculations

In order to produce the temperature profiles shown in Figure 3 above, we time evolved a rectangular 2D region on the silicon sample using the heat diffusion equation. The initial temperature versus depth profiles for this region were derived from the measured SIMS ion concentrations versus depths (Figure 3 (top right)). The ion concentrations were converted to temperature via the following scaling

$$\vec{T}_0 = \frac{ions}{cm^3} \cdot \frac{1}{C_p \rho} \cdot \vec{E}$$

where $\vec{E}$ is an array of ion energies (interpolated from the SIMS depth via a SRIM table for carbon ions in silicon) and $\vec{T}_0$ is a vector of initial temperatures over the spot depth. As reported above, this conversion produced an unrealistically high local temperature, which — when extrapolated over the entirety of the sample irradiated by the ion pulse — would approach or exceed the 35 J total energy of the laser pulse. We consider two possible explanations for the enhanced carbon concentration at shallow depth: 1) The measured temperature enhancement results from low energy carbon ions and particles from plasma expansion that reach a pre-heated surface and then rapidly diffusing into the sample in a manner similar to the diffusion of oxygen from the laser annealing of silicon with ~150–200 ns laser pulses [1]. 2) Since each SIMS spot only covers 100 μm², we may only be observing some small regions where the sample becomes exceedingly hot while others are much colder. This effect would be attributed to the fact that the beam is filamented and comprised of multiple high-intensity jets of ions. Hence, extrapolating locally very high energy deposition values from our SIMS measurement to the entire sample would be an overestimation. Considering that the values for energy deposition vary across the sample and can be very high in selected areas, the overall energy in the ion pulse is consistent with the TNSA mechanism for laser to ion energy conversion in the few % range [2].

Based on the many SIMS profiles at different locations within the footprint of the beam with high concentrations at shallow depth and the resemblance of our enhancement to the diffusion of oxygen and carbon during laser annealing, we take the diffusion hypothesis to be plausible and have allowed for this in the analysis of the SIMS depth profiles. Based on the diffusivity of carbon and other elements in solid silicon, we assume the silicon must be at least above the melting temperature for substantial diffusion to occur on the relevant timescale. We calculated the duration the sample must remain above either the boiling or melting point in order to create the diffusion of at least 1 µm into the surface (Figure 3 (top)).

$$\Delta t_{melt}(1\mu m) \simeq \frac{L^2}{D_{melt}} = \frac{(1 \cdot 10^{-4} cm^2)^2}{3.5 \cdot 10^{-5} \frac{cm^2}{s}} = 3 \cdot 10^{-4} s$$

$$\Delta t_{boil}(1\mu m) \simeq \frac{L^2}{D_{boil}} = \frac{(1 \cdot 10^{-4} cm^2)^2}{2 \cdot 10^{-4} \frac{cm^2}{s}} = 5 \cdot 10^{-5} s$$

where $D_{melt}$ and $D_{boil}$ are diffusion coefficients extrapolated from measurements in Scharmann *et al.* and Narayan *et al.*, respectively [3][4]. Constrained by these published values for the diffusivity of carbon in silicon, both $\Delta t_{melt}$ and $\Delta t_{boil}$ are significantly longer than the 0.1 µs for which the sample is above the boiling points and the 1 µs for which the sample is above the melting point according to the finite-element heat diffusion simulation (Figure 3), suggesting that there would not be enough time for surface carbons to diffuse 1 µm into the sample before it cools to below the melting point. Caveats to this suggestion include that the temperatures in Figure 3 are lower-limit estimates and that the diffusion coefficients might be extrapolated beyond their range of validity in order to calculate these estimates.

Notwithstanding the caveats and timescale issue above, we have calculated temperature profiles from the SIMS data by attributing the shallow-depth enhancement to diffusion of carbon from the surface. We therefore ignore the peaked diffusion component of the SIMS spectrum and make an exponential extrapolation to shallower depths based on the distribution for depths greater than ~2 µm (blue squares in Figure 3). The combination of the SIMS distribution at >2 µm and the extrapolation to shallow depth is taken to be the initial ion concentration profile in our 2D sample-heating model. The ion concentration profile is used to calculate an initial temperature profile.

Our simulation covers a region of 70 µm depth and 6 mm transverse width on our silicon sample. An initial gaussian transverse profile is assumed for the ion beam pulse (σ = 1.7 mm based on our NRA measurement) and the temperature vs depth at the center of the ion pulse is normalized to the SIMS-derived temperature profile. The gaussian is truncated at r = 0.5 mm to account for the ~1 mm beam diameter corresponding to the ~1 mm wide area where we observe surface exfoliation. Outside the 1 mm spot the temperature is set initially to 293 K. In order to simulate heat diffusion in the region over time, we solve the heat diffusion equation in the manner of Incropera *et al.* [5] (5.72):

$$\frac{1}{\alpha}\frac{\partial T}{\partial t} = \frac{\partial^2 T}{\partial x^2} + \frac{\partial^2 T}{\partial y^2}$$

where $\alpha = \frac{\kappa}{C_p \rho}$ is the thermal diffusivity. Similar to Table 5.3 in Incropera *et al.*, we produce the following three finite-difference equations for interior, side, and corner nodes, respectively, in our region of interest under adiabatic conditions

$$T_{m,n}^{p+1} = \frac{\alpha \Delta t}{\Delta x^2}[T_{m+1,n}^p + T_{m-1,n}^p] + \frac{\alpha \Delta t}{\Delta y^2}[T_{m,n+1}^p + T_{m,n-1}^p] + [1 - \frac{2\alpha \Delta t}{\Delta x^2} - \frac{2\alpha \Delta t}{\Delta y^2}]T_{m,n}^p$$

$$T_{m,n}^{p+1} = \frac{2\alpha \Delta t}{\Delta x^2}T_{m-1,n}^p + \frac{\alpha \Delta t}{\Delta y^2}[T_{m,n+1}^p + T_{m,n-1}^p] + [1 - \frac{2\alpha \Delta t}{\Delta x^2} - \frac{2\alpha \Delta t}{\Delta y^2}]T_{m,n}^p$$

$$T_{m,n}^{p+1} = \frac{2\alpha \Delta t}{\Delta x^2}T_{m-1,n}^p + \frac{2\alpha \Delta t}{\Delta y^2}T_{m,n-1}^p + [1 - \frac{2\alpha \Delta t}{\Delta x^2} - \frac{2\alpha \Delta t}{\Delta y^2}]T_{m,n}^p$$

where $T_{m,n}^p$ is the temperature of the node at the grid location (m, n) at the pth time step, $\Delta t$ is the size of the time step, and $\Delta x$ and $\Delta y$ are the grid step sizes in x and y, respectively. Using a Python script, we recursively solve these equations to construct temperature profiles at times as early as 1 ns and as late as 1 ms after the initial heating of the sample. The depth profiles in Figure 3 were constructed by extracting the temperatures at the center of the beam pulse at the times of interest. The Python code was benchmarked against the commercial heat-transfer finite difference program FEHT and was found to be in good agreement [6].

In summary, our observed shallow-depth (top few µm) enhanced carbon concentration qualitatively resembles that of a diffusion process driven by sudden heating of silicon samples via laser annealing with short-pulse lasers [1]. We modeled the heat diffusion of the ion energy deposition and calculated the characteristic time that the sample remains in the molten phase where the diffusivity of carbon in silicon is higher. The difficulty with this interpretation is that the estimated time for the carbon to diffuse to a depth of ~1 µm is longer than the cooling time. Yet, this depends greatly on the diffusion coefficients — for which there is considerable range in the literature — and the thermal conductivity of the samples [7].

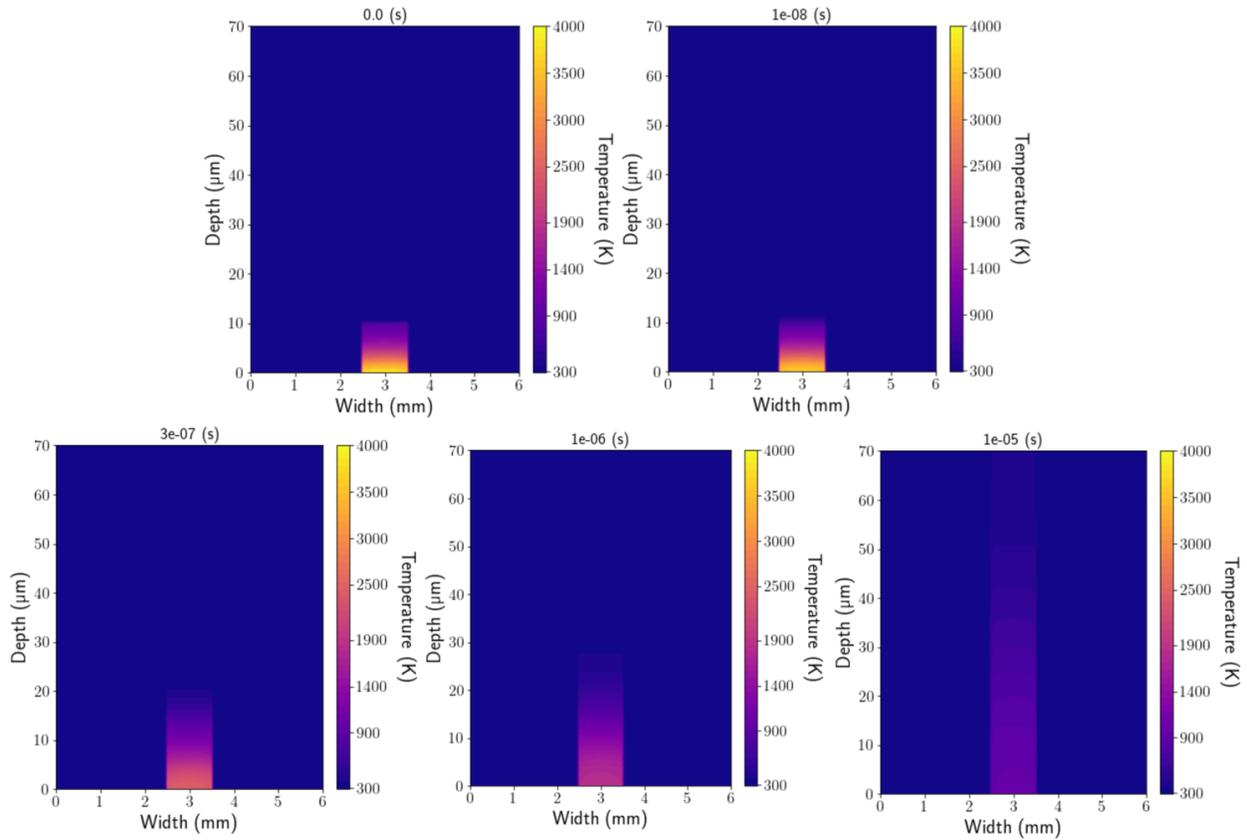

Figure S2: Frames of temperature evolution starting after the end of a 10 ns ion pulse to 10 μs. The sample cools to below the melting point after about 1 μs. The depth in this window is 70 μm and the beam spot is 1 mm wide in this example. Animation online: https://youtu.be/Su1zf5hLi8c .

3. NRA

Silicon samples that had been exposed to 10 and 100 ion pulses were analyzed using NRA for the detection and depth profiling of $^{12}C$, $^{14}N$ and $^{16}O$ isotopes through the $^{12}C(d,p_0)^{13}C$, $^{14}N(d,a_1)^{12}C$ and $^{16}O(d,p_1)^{17}O$ nuclear reactions [8,9], respectively. The measurements were carried out with a 1.4 MeV deuteron beam, a beam diameter of 1 mm and a Passivated Implanted Planar Silicon (PIPS) Detector with an area of 300 mm$^2$ placed at the scattering angle θ=150º. A 13 μm thickness filter of aluminized Mylar was placed in front of the PIPS to avoid that elastically scattered deuterons reach the detector, which could saturate the electronic chain and produce pile-up. Reference samples consisting of thin layers of $Si_3N_4$/Si and $Ta_2O_5$/Ta with a known amount (±3%) of N and O were used to calibrate the NRA spectra, which were analyzed using the SIMNRA 6.0 code [8].

Figure S3 shows an image of the surface of two samples after 10 and 100 shots, respectively, where it is apparent that the irradiated area was very inhomogeneous. A scan of 6 (10x sample) and 8 (100x sample) NRA measurements along the horizontal direction was performed with 1 mm steps/point. An ion beam dose of 5 µC was accumulated for every point of analysis.

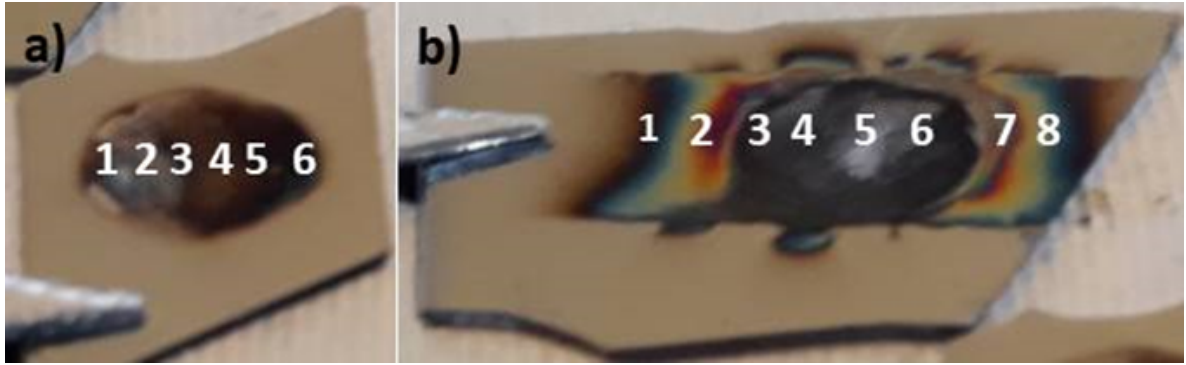

Figure S3: Photograph of the samples after (a) 10 ion shots and (b) 100 ion shots with the corresponding impact points where the NRA measurements have been performed. The measurement beam spot diameter was 1 mm at each measurement position.

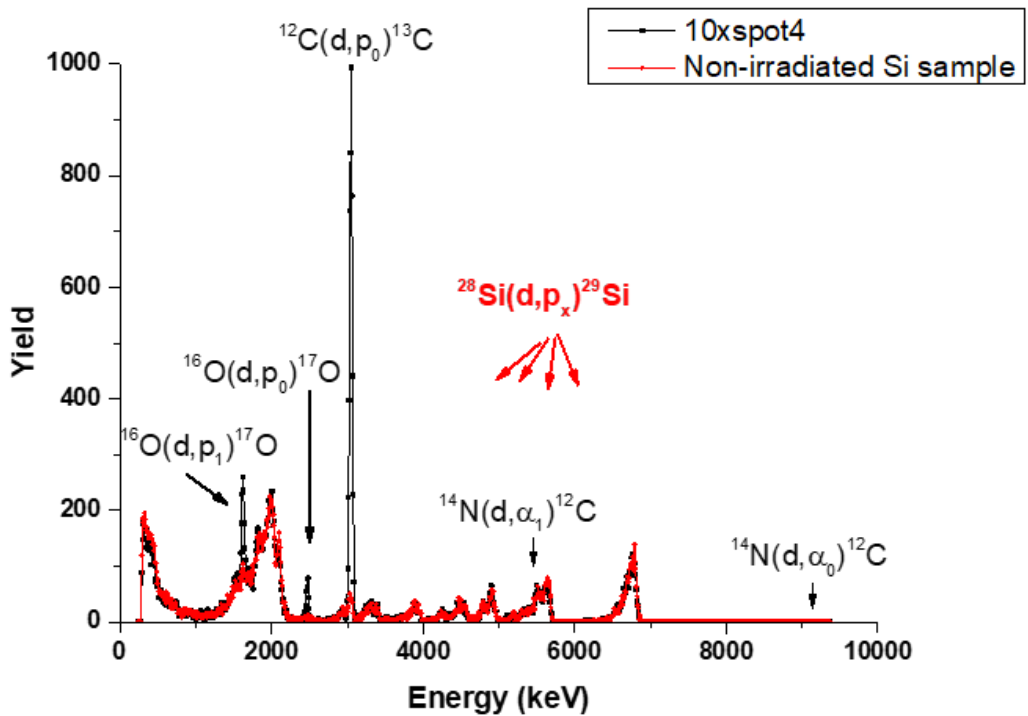

Figure S4: Example of NRA spectra measured in an irradiated point (black signal) and in a pristine Si sample (red signal).

Figure S4 shows the NRA spectra obtained in an irradiated point (black spectrum) in comparison with the spectrum from an as received silicon sample (red spectrum). Except for the most intense signal, corresponding to the $^{14}$N(d,a$_0$)$^{12}$C reaction, it is evident that there is strong overlap of the peaks coming from the $^{12}$C, $^{14}$N and $^{16}$O isotopes with the signals produced with the $^{28}$Si from the substrate that populate various excited states of $^{29}$Si nuclei. Due to the unavailability of several cross sections data for the $^{28}$Si(d,p$_x$)$^{29}$Si nuclear reactions, it was not possible to make a global fit of the spectra with the SIMNRA program. To overcome this difficulty and to be able to carry out the simulations, the spectrum from a pristine silicon target was subtracted from all measured spectra.

Representative spectra obtained at different positions of the sample irradiated with 10 ion pulses, after subtraction of the background signal, are shown in Figure S5. For all measured points we observe signals from carbon and oxygen atoms in the top 200 nm near the surface with respective fluences ranging between 75–250×10$^{15}$ C/cm$^2$ and 75–125×10$^{15}$ O/cm$^2$. Only point 2 shows a depth profile for carbon, adding a supplementary fluence of 30×10$^{15}$ C/cm$^2$, which extends to a depth of ~ 3.2 μm. For oxygen, it is not possible to know if the few counts that appear at energies lower than 1600 keV are due to a real oxygen profile or to an imperfect subtraction process of the Si signal. The nitrogen peaks (not shown) are very small and correspond to fluences between 9-27×10$^{15}$ N/cm$^2$.

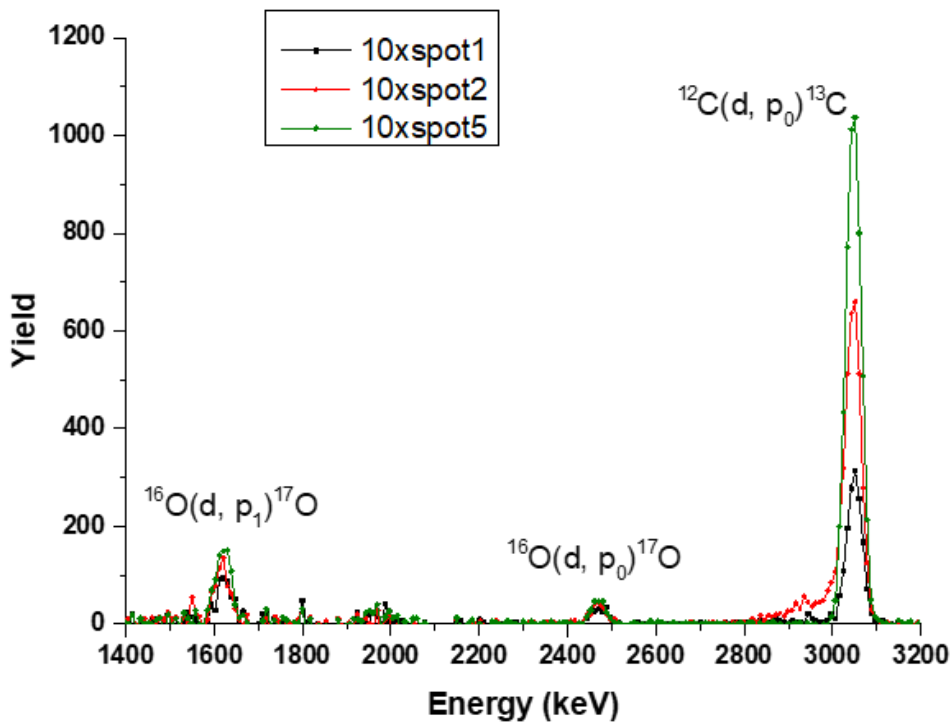

Figure S5: NRA spectra after subtraction of the Si substrate signal for a series of positions across the 10x pulse sample corresponding to the areas in the image in Figure S3 (a).

The NRA spectra acquired across a series of positions of the 100x pulse sample, together with the SIMNRA simulation of the spectrum obtained at point 5, are displayed in Figure S5. The resulting carbon profiles extracted from the NRA measurements are depicted in Figure 3. As expected from the optical image, the impurity content is very inhomogeneous, with values in the first 200 nm ranging between 245–1100×10$^{15}$ C/cm$^2$, 75–260×10$^{15}$ O/cm$^2$ and 15–89×10$^{15}$ N/cm$^2$, respectively.

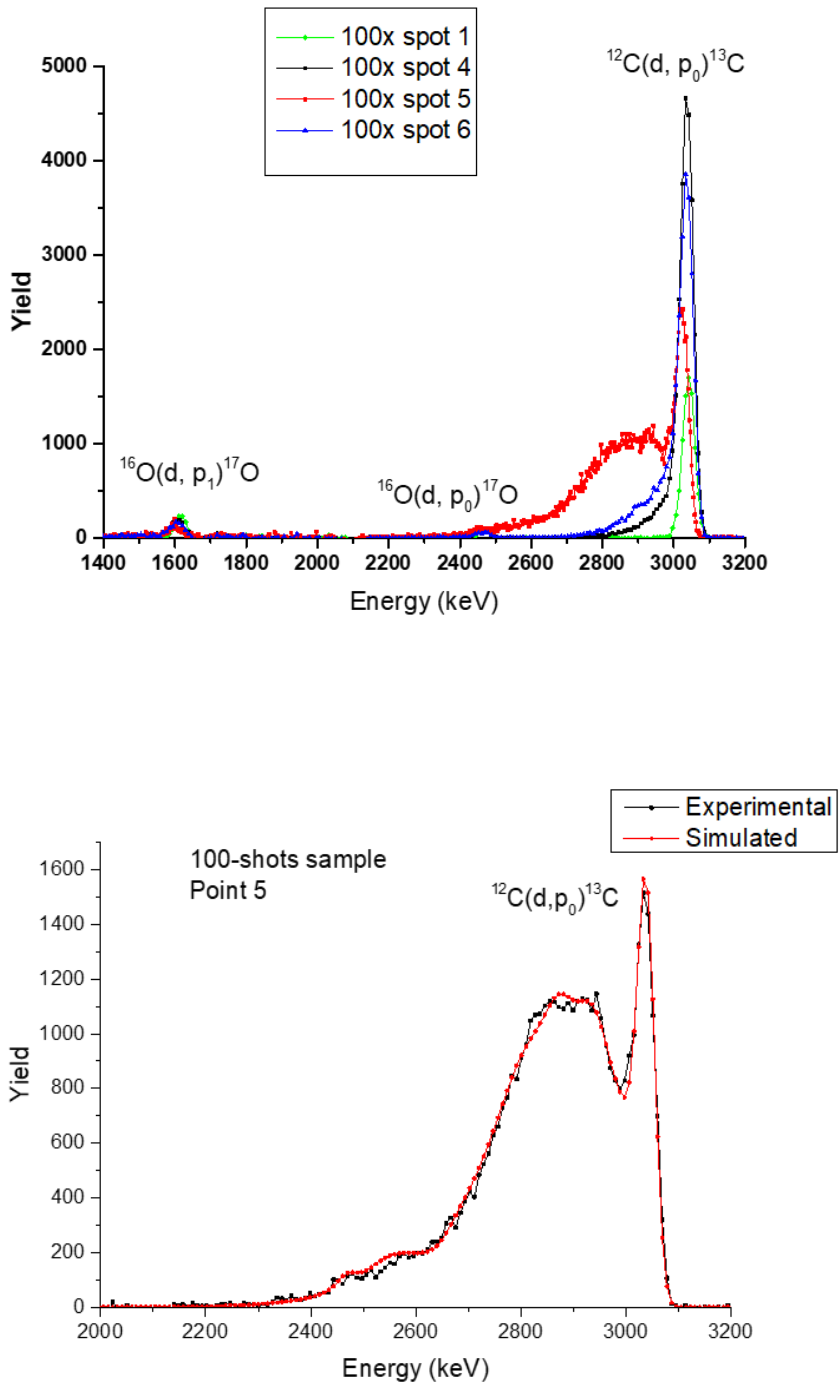

Figure S6: Top: NRA spectra after subtraction of the Si signal for a series of positions across the 100x pulse sample corresponding to the areas in the image in Figure S3 (b). Bottom: Experimental and SIMNRA simulated spectra of the $^{12}$C(d,p$_0$)$^{13}$C signal from spot 5.

For this sample, the deepest carbon atoms are highly localized (only at point 5) and they reach a depth of ~ 9.2 µm.  As discussed in the main text, a high fluence of low energy carbon ions, ~$10^{16}$ cm$^{-2}$ per pulse is implanted near the surface of the samples.  Carbon at depths > 1 µm is from in-diffusion of carbon with a contribution of high energy ions from TNSA.

4. **Channeling RBS**

In order to quantify the structure of the silicon (111) crystals after a series of laser ion pulses, we conducted Rutherford Backscattering analysis in channeling direction (ch-RBS)[9,10].  We used 2 MeV helium ions with a probe beam spot of 1 mm diameter.  The energy resolution of the detector was 18 keV.  The analysis geometry is shown in Figure S4 together with typical spectra for silicon (111) samples that had been exposed to 1, 2, 10 and 100x laser ion pulses.  Ch-RBS probes the top ~2 µm of the samples.

Ch-RBS is sensitive to the accumulation of radiation damage, such as point defects and dislocation loops, as well as surface structure changes from the impact of energetic ions on silicon crystals.  We observe increasing backscattering yields in the channeling direction, increasing from the familiar very low yields in the pristine material for samples that received increasing numbers of pulses (Figure S5).  But the total accumulated disorder is much lower than expected for carbon ion irradiation with an energy deposition in the ~2 J/cm$^2$ range and carbon ion energies ranging from 0.1 to 8 MeV with fluences as shown in Figure 3 in the main text [10–12].  This is consistent with the high thermal budget from exposure to intense laser-ion pulses leading to point defect annealing and dislocation loop formation.

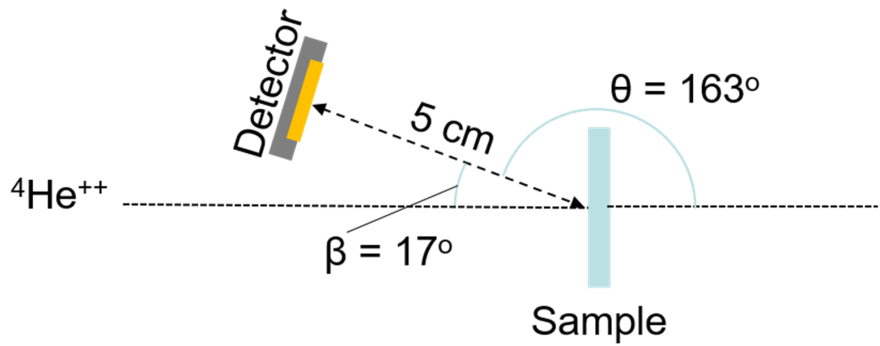

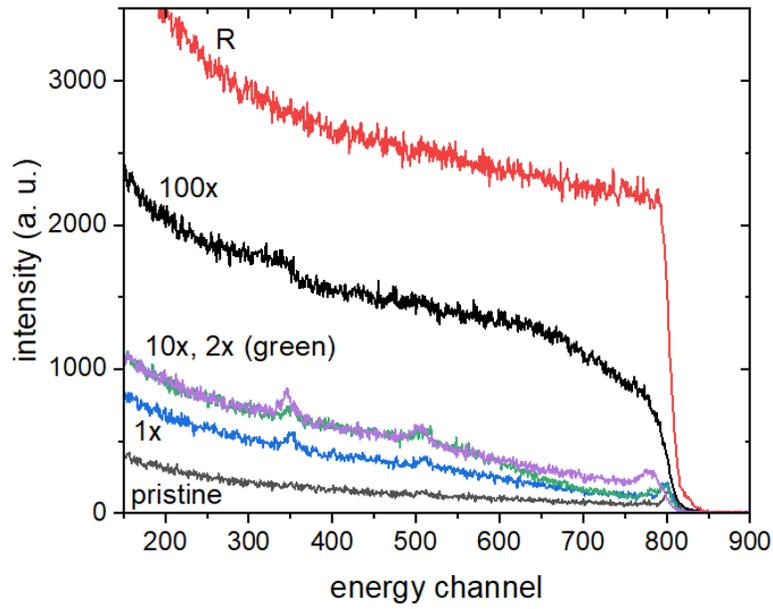

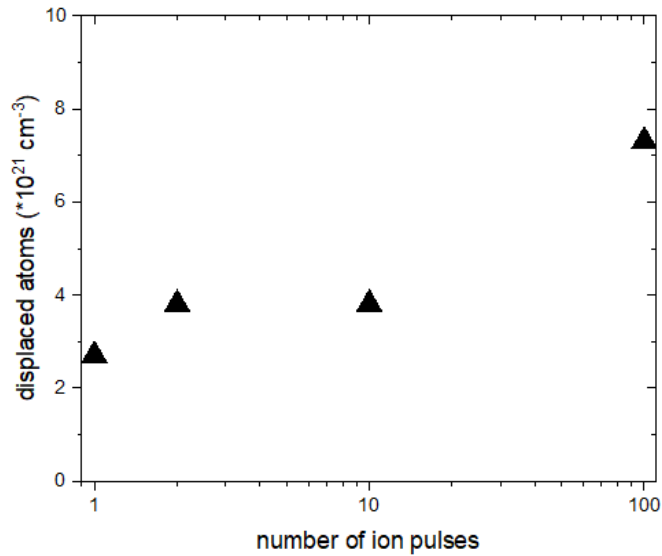

Figure S7: Top: Schematic of the ch-RBS setup. Mid: Ch-RBS spectra for silicon (111) samples that had been exposed to 1, 2, 10 and 100x ion pulses, compared to a spectrum taken with random alignment (R, red). Bottom: Concentration of displaced silicon atoms as a function of the number of laser ion pulses a silicon (111) had received.

## 5. Photoluminescence spectra of photon emitting defects and SIMS depth profiles at the same locations

In Figure S8 we complement the PL results from Figure 5 in the main text with SIMS depth profiles of carbon, oxygen and hydrogen of a sample that had received two laser-ion pulses. The SIMS depth profiles were taken in the same areas on the sample where PL spectra had been taken earlier. We observe G-centers from proton irradiation in areas just outside the central beam spot area that had been covered with aluminum foil (left). In areas with increased carbon concentrations we see W-centers with a narrow linewidth distribution and G-centers with a broadened distribution. Areas with the highest thermal budget and ion flux show W-center ensembles with a slightly broadened linewidth. The shape of the SIMS profile from the high flux area on the right is indicative of surface roughness increases due to the onset of exfoliation.

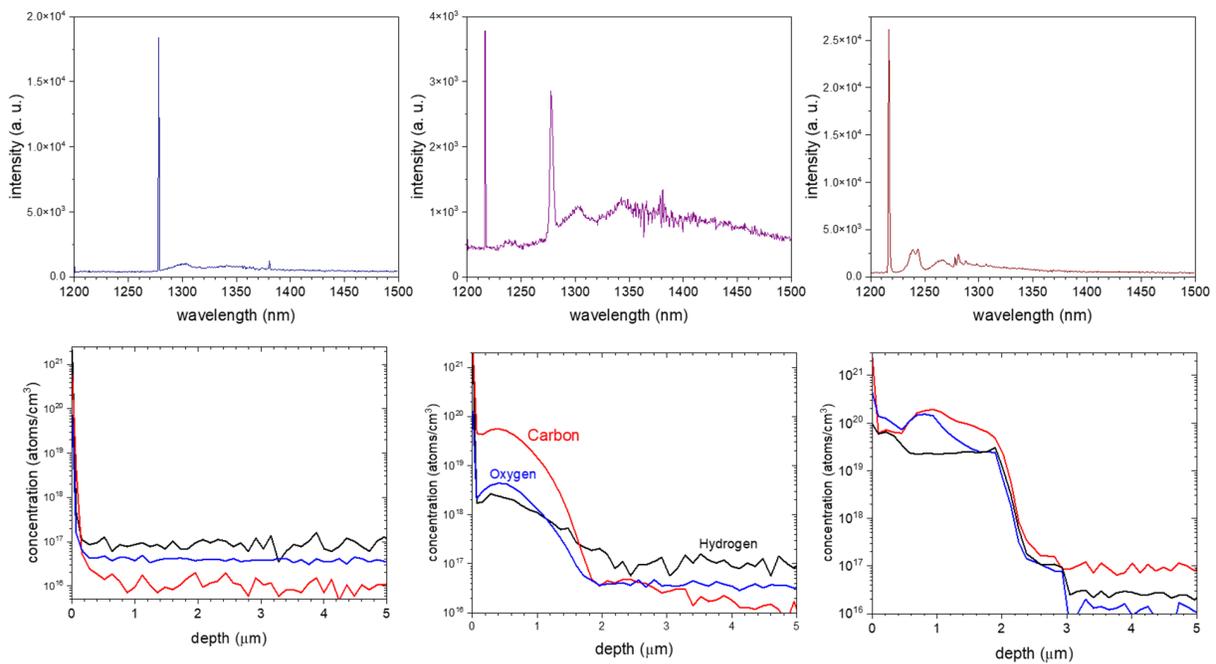

Figure S6: PL and SIMS results from a silicon (111) sample after two laser ion pulses. The PL spectra were taken first and SIMS depth profiles were then taken afterwards in the same locations. Left: Narrow G-centers in areas with proton irradiation only (covered with Al foil), mid: narrow W and broadened G-center from areas with increase carbon drive in diffusion and MeV carbon and proton flux, right: slightly broadened W-centers in areas with the highest energy deposition from laser ion pulses.

## 6. Simulations of gaps between defect levels for the G-center and W-center

To understand the effect of excess carbon in proximity to a photon emitting defect, we performed calculations on the structures of G and W-centers embedded within a 3×3×3 silicon unit cell, with an additional interstitial carbon placed at one of the 106 possible tetrahedral sites within the supercell. The structures are fully relaxed and the energy splitting between the two defect levels within the silicon band gap are plotted vs the distance between the defect and an interstitial carbon atom. When the carbon interstitial is close to the defect (< 5 Å), there is a dramatic effect on the defect levels. Greater separations (> 10 Å) correspond more closely to the expected separation between defects and stray interstitials at the typical observed concentrations below a few atomic % carbon in silicon. For these configurations, the G-center still has some spread in the gap energies of different configurations, but for the W-center the gaps are nearly identical, in line with the notion that the G-center is much more sensitive to local disorder than the W-center.

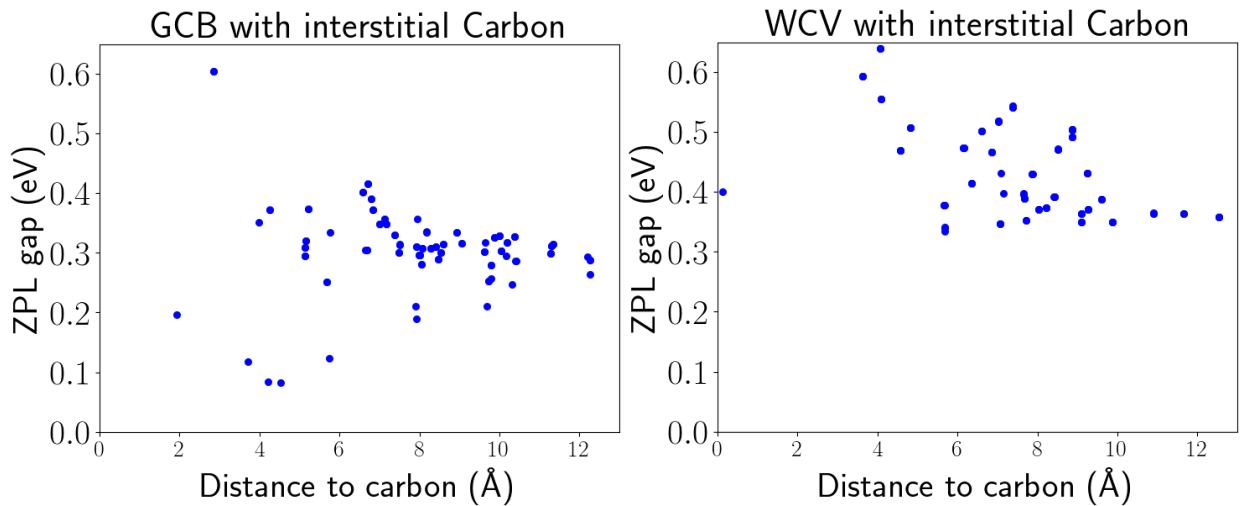

Figure S7: Zero-phonon line gap, calculated as the difference between lower and upper defect levels for G-centers (left) and W-centers (right) modulated by nearby interstitial carbon sites, plotted versus distance between the defect and interstitial carbon.